# Evaluating the Impact of Population Growth and Applying Decarbonization Regulations on the Energy Market in the ERCOT Interconnection Area

**Farah Altarazi, PhD**
**The State University of New York at Binghamton**
**New York, USA**

faltara1@binghamton.edu

### ERCOT Energy Background

ERCOT interconnection area covers one state, Texas. In 2022, Texas produced more electricity than any other state and generated twice as much as second-place Florida and accounted for more than 12% of the nation's total electricity net generation that year (U.S. Energy Information Administration, 2023). According to "Texas Electricity Profile 2022" which is published by the U.S. Energy Information Administration, the total net generation for the year 2022 is 525,562,940 Megawatt Hours (MWh) where natural gas is considered the primary energy source. The total retail sales of 475,401,192 MWh with an average retail price of 10.16 (Cents/kWh) (U.S. Energy Information Administration, 2023).

ERCOT in their most recent "ERCOT Fact Sheet 2023" set an all-time peak demand record of 85,464 MW on August 10, 2023, and 85,116 MW for weekend peak demand record on August 20, 2023 (ERCOT, 2023). Figure 1 shows a comparison between load duration curves for each year from 2020 through 2022 in ERCOT, the load over the years has almost the same shape and it is increased from one year to another (Economics, 2023).

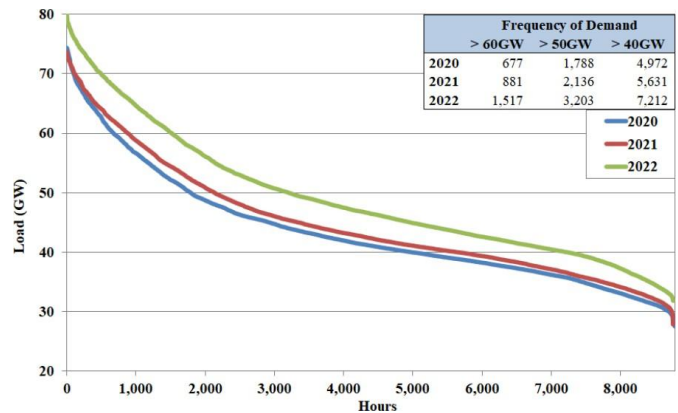

*Figure 1 Load Duration Curves for Each Year From 2020 Through 2022 (Economics, 2023)*

The annual generation mix for the years 2014 to 2022 is shown in Figure 2. For the year 2022, Natural gas is considered the main contributor and accounts for almost 42.5% of the total share, compared to 41.9% in 2021. The figure also shows that the generation share from wind has increased every year and accounts for 25% in the year 2022, the same thing applies to solar energy with a slightly less increased rate over the years and a total share of 5.6% in 2022. Another thing that can be noted is that the share of generation from coal dropped from 19.0% in 2021 to 16.6% in 2022 (Economics, 2023). According to the U.S. Energy Information Administration's latest data, in 2022, Texas generated 26% of all U.S. wind-sourced electricity, leading the nation for the 17th year in a row.

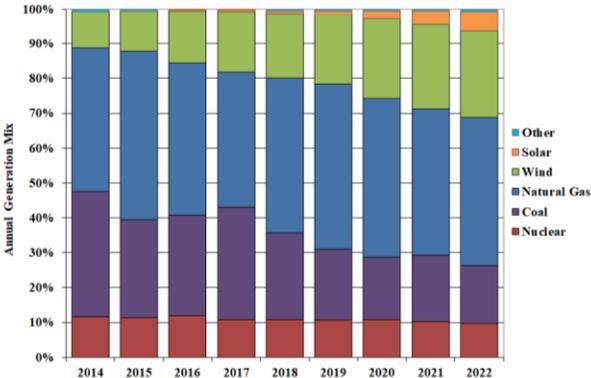
*Figure 2 Annual Generation Mix in ERCOT (Economics, 2023)*

Texas leads the nation in energy consumption across all sectors and is the largest energy-consuming state in the nation, Table 1 shows the comparison between the year 2021 and year 2022 electricity consumption per sector, the transportation sector has the highest number, followed by residential, then commercial, and last one is industrial (U.S. Energy Information Administration, 2023). Overall, the consumption of 2022 for all sectors increased compared with the year 2021.

*Table 1 Breakdown of Electricity Consumption into Sectors (Thousand Megawatt hours) (U.S. Energy Information Administration. ,2023)*

| Residential | | Commercial | | Industrial | | Transportation | |
|---|---|---|---|---|---|---|---|
| 2022 | 2021 | 2022 | 2021 | 2022 | 2021 | 2022 | 2021 |
| 170,596 | 155,075 | 160,719 | 147,843 | 143,906 | 132,530 | 180 | 180 |

**Research Questions & Motivation**

The World Energy Counsel mentioned in their recent World Energy Trilemma Index 2022 that energy decision-makers should organize the supply and manage the competing demands based on the three energy dimensions which are Energy Security, Energy Equity, and Environmental Sustainability to enhance their annual measurement of national energy system performances (The World Energy Counsel, 2022), based on that the energy providers must focus on supplying current and future energy demand reliably and to provide universal access to affordable, and abundant energy for domestic and commercial use, besides keeping in mind the transition of a country's energy system towards mitigating and avoiding potential environmental harm and climate change impact.

In order to be able to achieve these national and global goals, the factors that affect energy supply and demand should be studied and managed effectively. One of these factors is population growth, with the increased population growth throughout the states, Texas is considered one of the most affected ones that experienced steady population growth from 2010 till now, its population growth rate has dramatically increased, and it is expected to continue rising in the upcoming years (Potter& Hoque, 2014), as shown in Figure 3. This will make the process of organizing the energy market and managing electricity supply and demands to achieve a balance among the three energy dimensions more challenging.

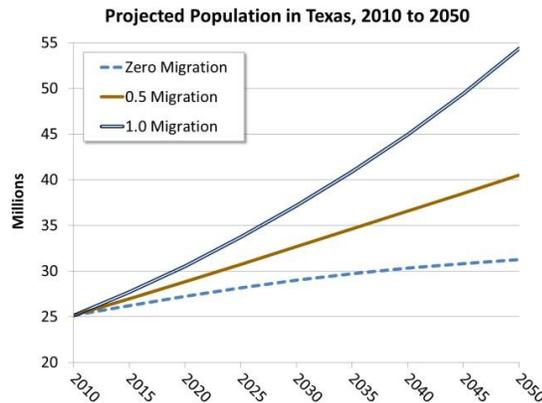
*Figure 3 Projected Population in Texas (2010-2050) (Potter& Hoque, 2014)*

Additionally, it is globally known that Gross Domestic Product (GDP) per capita and population growth stand out as the primary drivers of carbon dioxide ($CO_2$) emissions over the last decade (Energy Information Administration, 2023), Figure 4 shows a direct relation between $CO_2$ emissions and population growth over the years, in which as the population increased the $CO_2$ emissions also increased. It is known also that the largest source of greenhouse gas emissions from human activities in the United States is from burning fossil fuels for electricity, heat, and transportation, over 40% of energy-related $CO_2$ emissions are due to the burning of fossil fuels for electricity generation (Energy Information Administration, 2023), which also highlight the importance of studying the current decarbonization regulations in Texas over the next decade advocating for the using of less environmentally harmful resources.

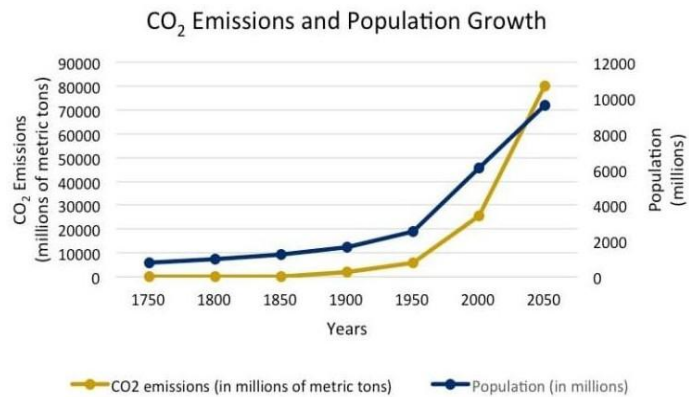
*Figure 4 Relation CO2 Emissions and Population Growth (Energy Information Administration, 2023)*

This study will contribute to enhancing the understanding of the above-mentioned effects and factors on the ERCOT energy market by addressing two research questions:

- How do the population growth and the increased demand impact the energy market in the ERCOT interconnection area?
- The increased population growth will result in increased $CO_2$ emissions, how can various regulations and reform markets promote the decarbonization of the energy system in the ERCOT interconnection area over the next decade?

**Methodology**
<u>Model Description</u>

We will use a complex capacity expansion model and a linear programming solver, CLP, to solve a constrained optimization problem to determine total costs, the mix of generators, non-served energy (NSE), storage resource capacity, and the total transfer energy to meet the demand in the planning year 2030, this will be done for different three zones in ERCOT interconnection area. The current formulation of the model runs with a 480-time index for the year 2030 (Neha Patankar, SR1).

Broadly, the model considers the investment costs of new resources, operational fixed, start-up, and variable costs of resources, operational constraints of unit commitment and storage resources, and the variability of renewable resources and load, as well as bulk transmission constraints between zones. For thermal generators, including nuclear, natural gas, and coal plants, which require integer unit commitment (UC) decisions (to start up or shut down), this model uses a linear relaxation of the UC constraints, which uses clusters of similar generators (e.g. based on heat rate) for the UC decisions. The implementation of linear relaxation for the UC constraints reduces the complexity of the optimization problem to a linear programming (LP) problem instead of a mixed integer linear programming (MILP) problem and reduces the computational solving time (Neha Patankar, SR1).

To address the research questions, we will first investigate the relationship between population growth and demand. Subsequently, we will use this relationship to analyze the effects of both an increase (50%) and a decrease (50%) in population on demand for the year 2030, the choice of 50% was based on the expectation that in 2030, population growth may encounter three distinct scenarios, each differing by a 50% migration rate from the others (Potter& Hoque, 2014). We will then investigate the associated changes in the ERCOT interconnection area energy market using the 3-zone Complex Capacity Expansion model.

Then to check the effect of applying decarbonization regulations on the energy market in the ERCOT interconnection area, two policies have been investigated for each of the population scenarios. In June 2014, the U.S. Environmental Protection Agency (EPA) proposed the Clean Power Plan, which calls for reductions in the carbon intensity of the electric sector. The Clean Power Plan would set limits on the carbon dioxide ($CO_2$) emissions from existing fossil fuel-fired power plants, calculated as state emissions rate goals. For Texas, the EPA has proposed an interim goal of 853 lb $CO_2$/MWh to be met on average during 2020-2029 and a final goal of 791 lb $CO_2$/MWh to be met from 2030 onward (ERCOT Public, 2015).

However, these values have been adjusted, EPA released details of the CPP final rule on August 3, 2015. In the final rule, several changes were made to the proposal, including modifications to the emissions limit calculation and the compliance deadlines. Under the CPP final rule, Texas will be required to meet a final $CO_2$ emissions rate limit of 1,042 lb $CO_2$/MWh (0.521 t$CO_2$/MWh) on average from 2030 onwards, or 190 million tons of $CO_2$. EPA calculated these limits based on assumptions about coal plant efficiency improvements, increased production from natural gas combined cycle units, and growth in generation from renewable resources. EPA also modified the compliance deadlines in the final rule, phasing in the reductions over three interim compliance periods between 2022 and 2029 (ERCOT Public, 2015).

Another policy to reduce carbon emissions is the carbon pricing technique. A price of $22.50 per ton of $CO_2$ is implemented. This pricing strategy has been suggested in conjunction with the CPP (Clean Power Plan) and the Cross-State Air Pollution Rule (CSAPR) (ERCOT Public, 2015).

Input Data

To build the above model we will use the available ERCOT data (ERCOT_4x5 dataset) with a 480-time index, that includes datasets with the required information for the generators, demand, generators variability, fuels, network, zones, and lines. The dataset includes information on time intervals, with four representative periods (Rep_Periods) identified. Within each period, there are varying numbers of time steps (Timesteps_per_Rep_Period), ranging from 120 to 3120. Additionally, Sub_Weights are assigned to different components for each period.

*Population input data*

To find the effect of population, we need to connect the population estimates with the electricity demand estimates. Power consumption hit its most recent record of 85,435 megawatts on Aug. 10, 2023, according to ERCOT (ERCOT, 2023). The primary drivers of record consumption have been the rapid growth in the state's resident population and economy, power consumption increased at a compound annual rate of 1.7% between 2001 and 2022 – broadly in line with the compound population increase of 1.6% annually (Kemp, 2023). We will use this rate to connect the population and the load in our dataset for the year 2030.

The statement suggests a correlation between power consumption and population increase over the period from 2001 to 2022:

- Power Consumption Increase Rate: The power consumption increased at a compound annual rate of 1.7%.
- Population Increase Rate: The population increased at a compound annual rate of 1.6%.
- The rate of increase in power consumption is roughly proportional to the rate of population growth.

Figure 5 illustrates the approximate correlation between population and electricity consumption from the years 2001 to 2022. The graph reveals a direct linear relationship with a high R-square value of 0.9637, indicating a strong correlation (Durden, 2023).

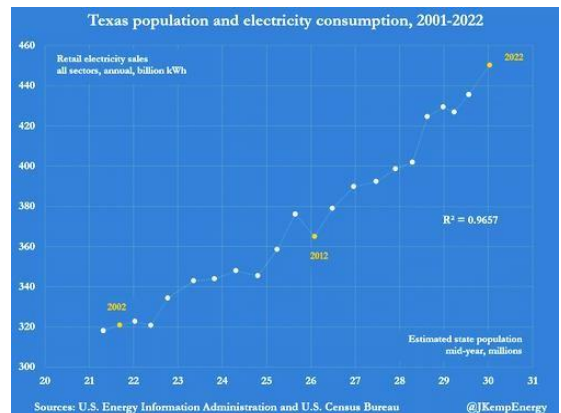

*Figure 5 Correlation Between Power Consumption and Population in Texas during years (2001-2022) (Durden, 2023)*

Based on that, we will find K, which represents the proportionality constant in the relationship between the rate of power consumption growth (Equation 1) and the rate of population growth (Equation 2). In this context, k is calculated as the ratio of the compound annual growth rate (CAGR) for power consumption (0.017) to the CAGR of population growth (0.016) (Eqaution3):

*Power Consumption = Initial Power Consumption × (1 + CAGR Power Consumption) ^ time*   (1)

*Population Consumption = Initial Population ×(1+CAGR Population)^time*     (2)
*Dividing the two equations:*
*k = CAGR Power Consumption / CAGR Population*     (3)

This implies that the rate of power consumption growth is approximately 1.0625 times the rate of population growth over the specified number of years.

Model Assumptions
- We will assume that (K) applies to the population and demand growth for the year 2030.
- We will assume that the considered period represents the whole year of 2030.

Scenarios Development

To answer the research questions, different scenarios have been developed that link the cases of changing population growth rate and applying decarbonization policies, which results in nine different scenarios, as shown in Figure 6.

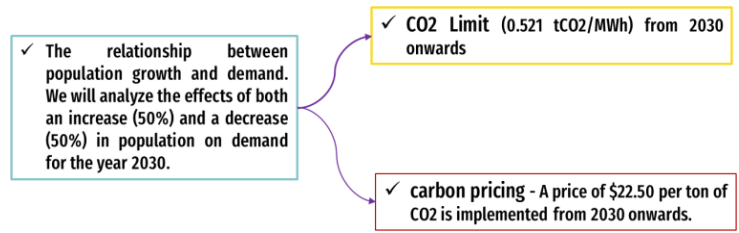

Figure 6 Different Research Scenarios

Models Developments
1. Models scenarios with changing population growth rates

These are the three models that will show the effect of changing population growth rates, as shown below:
  a) Basic scenario- we will use the provided load data with no changes.
  b) 50% decreased population scenario- Load (MW) in case of population growth less than 50% from the basic scenario.
  c) 50% increased population scenario- Load (MW) in case of population growth of more than 50% from the basic scenario.

To build these models we will develop complex capacity expansion models and a linear programming solver, CLP, with the defining the following sets, parameters, decision variables, constraints, and objective function ((Neha Patankar, SR1).

- Sets

G: Set of all generators, S: Set of all non-served energy (price responsive demand) segments, P: Set of time sample sub-periods (e.g. sample days or weeks), W: Sub-period cluster weights = number of periods (days/weeks) represented by each sample period, T: Set of all-time steps, Z: Set of zones (by number), L: Set of all lines

- Parameters

nse: non-served energy parameters (by s in S), generators: generation (and storage) parameters (by g in G), demand: demand parameters (by t in T), zones: network zones (by z in Z), lines: transmission lines (by l in L)

- Subsets

UC: Subset of G of all thermal resources subject to unit commitment constraints, ED: Subset of G NOT subject to unit commitment constraints, STOR: Subset of G of all storage resources, VRE: Subset of G of all variable renewable resources, NEW: Subset of all new build resources, OLD: Subset of all existing resources, RPS: Subset of all RPS qualifying resources, For the ERCOT area, there are no RPS policies, where *RPS* denotes the set of RPS-eligible resources in Texas State, including solar thermal, solar photovoltaic, wind, biomass, hydroelectric, combined heat and power, fuel cells, landfill gas, tidal, wave, ocean thermal, and anerobic digestion.

- Decision variables

The decision variables of the model are the variables that represent decisions in capacity additions, denoted as *capacity decision variables*, and decisions regarding electricity generation and resource operation, denoted as *operational decision variables*. All decision variables in this model are continuous numbers and are all greater than or equal to zero, except for transmission flow, $FLOW_{h,\,l}$, which can also take negative values depending on the direction of electricity flow.

Capacity decision variables

```
vCAP[g in G]                         >= 0    # power capacity (MW)
vRET_CAP[g in OLD]                   >= 0    # retirement of power capacity (MW)
vNEW_CAP[g in NEW]                   >= 0    # new build power capacity (MW)

vE_CAP[g in STOR]                    >= 0    # storage energy capacity (MWh)
vRET_E_CAP[g in intersect(STOR, OLD)] >= 0   # retirement of storage energy capacity (MWh)
vNEW_E_CAP[g in intersect(STOR, NEW)] >= 0   # new build storage energy capacity (MWh)

vT_CAP[l in L]                       >= 0    # transmission capacity (MW)
vRET_T_CAP[l in L]                   >= 0    # retirement of transmission capacity (MW)
vNEW_T_CAP[l in L]                   >= 0    # new build transmission capacity (MW)
```

Operational decision variables

```
vGEN[T,G]        >= 0  # Power generation (MW)
vCHARGE[T,STOR]  >= 0  # Power charging (MW)
vSOC[T,STOR]     >= 0  # Energy storage state of charge (MWh)
vNSE[T,S,Z]      >= 0  # Non-served energy/demand curtailment (MW)
vFLOW[T,L]             # Transmission line flow (MW);
  # note line flow is positive if flowing
  # from source node (indicated by 1 in zone column for that line)
  # to sink node (indicated by -1 in zone column for that line);
  # flow is negative if flowing from sink to source.
```

- Objective function

The objective function is to minimize the sum of fixed costs associated with capacity decisions and variable costs associated with operational decisions.

```julia
# Create expressions for each sub-component of the total cost (for later retrieval)
@expression(Expansion_Model, eFixedCostsGeneration,
    # Fixed costs for total capacity
    sum(generators.Fixed_OM_Cost_per_MWyr[g]*vCAP[g] for g in G) +
    # Investment cost for new capacity
    sum(generators.Inv_Cost_per_MWhyr[g]*vNEW_CAP[g] for g in NEW)
)
@expression(Expansion_Model, eFixedCostsStorage,
    # Fixed costs for total storage energy capacity
    sum(generators.Fixed_OM_Cost_per_MWhyr[g]*vE_CAP[g] for g in STOR) +
    # Investment costs for new storage energy capacity
    sum(generators.Inv_Cost_per_MWhyr[g]*vNEW_E_CAP[g] for g in intersect(STOR, NEW))
)
@expression(Expansion_Model, eFixedCostsTransmission,
    # Investment and fixed O&M costs for transmission lines
    sum(lines.Line_Fixed_Cost_per_MW_yr[l]*vT_CAP[l] +
        lines.Line_Reinforcement_Cost_per_MWyr[l]*vNEW_T_CAP[l] for l in L)
)
@expression(Expansion_Model, eVariableCosts,
    # Variable costs for generation, weighted by hourly sample weight
    sum(sample_weight[t]*generators.Var_Cost[g]*vGEN[t,g] for t in T, g in G)
)
@expression(Expansion_Model, eNSECosts,
    # Non-served energy costs
    sum(sample_weight[t]*nse.NSE_Cost[s]*vNSE[t,s,z] for t in T, s in S, z in Z)
)

@objective(Expansion_Model, Min,
    eFixedCostsGeneration + eFixedCostsStorage + eFixedCostsTransmission +
    eVariableCosts + eNSECosts
);
```

- Constraints

```julia
# (1) Supply-demand balance constraint for all time steps and zones
@constraint(Expansion_Model, cDemandBalance[t in T, z in Z],
        sum(vGEN[t,g] for g in intersect(generators[generators.Zone.==z,:R_ID],G)) +
        sum(vNSE[t,s,z] for s in S) -
        sum(vCHARGE[t,g] for g in intersect(generators[generators.Zone.==z,:R_ID],STOR)) -
        demand[t,z] -
        sum(lines[l,Symbol(string("z",z))] * vFLOW[t,l] for l in L) == 0
);
# Notes:
# 1. intersect(generators[generators.zone.==z,:R_ID],G) is the subset of all
# generators/storage located at zone z in Z.
# 2. sum(lines[l,Symbol(string("z",z))].*FLOW[l,t], l in L) is the net sum of
# all flows out of zone z (net exports)
# 3. We use Symbol(string("z",z)) to convert the numerical reference to z in Z
# to a Symbol in set {:z1, :z2, :z3} as this is the reference to the columns
# in the lines data for zone z indicating which whether z is a source or sink
# for each line l in L.

# (2-6) Capacitated constraints:
@constraints(Expansion_Model, begin
# (2) Max power constraints for all time steps and all generators/storage
    cMaxPower[t in T, g in G], vGEN[t,g] <= variability[t,g]*vCAP[g]
# (3) Max charge constraints for all time steps and all storage resources
    cMaxCharge[t in T, g in STOR], vCHARGE[t,g] <= vCAP[g]
# (4) Max state of charge constraints for all time steps and all storage resources
    cMaxSOC[t in T, g in STOR], vSOC[t,g] <= vE_CAP[g]
# (5) Max non-served energy constraints for all time steps and all segments and all zones
    cMaxNSE[t in T, s in S, z in Z], vNSE[t,s,z] <= nse.NSE_Max[s]*demand[t,z]
# (6a) Max flow constraints for all time steps and all lines
    cMaxFlow[t in T, l in L], vFLOW[t,l] <= vT_CAP[l]
# (6b) Min flow constraints for all time steps and all lines
    cMinFlow[t in T, l in L], vFLOW[t,l] >= -vT_CAP[l]
end)
```

```julia
# (7-9) Total capacity constraints:
@constraints(Expansion_Model, begin
# (7a) Total capacity for existing units
    cCapOld[g in OLD], vCAP[g] == generators.Existing_Cap_MW[g] - vRET_CAP[g]
# (7b) Total capacity for new units
    cCapNew[g in NEW], vCAP[g] == vNEW_CAP[g]

# (8a) Total energy storage capacity for existing units
    cCapEnergyOld[g in intersect(STOR, OLD)],
        vE_CAP[g] == generators.Existing_Cap_MWh[g] - vRET_E_CAP[g]
# (8b) Total energy storage capacity for existing units
    cCapEnergyNew[g in intersect(STOR, NEW)],
        vE_CAP[g] == vNEW_E_CAP[g]

# (9) Total transmission capacity
    cTransCap[l in L], vT_CAP[l] == lines.Line_Max_Flow_MW[l] - vRET_T_CAP[l] + vNEW_T_CAP[l]
end)

# (10-12) Time coupling constraints
@constraints(Expansion_Model, begin
    # (10a) Ramp up constraints, normal
    cRampUp[t in INTERIORS, g in G],
        vGEN[t,g] - vGEN[t-1,g] <= generators.Ramp_Up_Percentage[g]*vCAP[g]
    # (10b) Ramp up constraints, sub-period wrapping
    cRampUpWrap[t in STARTS, g in G],
        vGEN[t,g] - vGEN[t+hours_per_period-1,g] <= generators.Ramp_Up_Percentage[g]*vCAP[g]

    # (11a) Ramp down, normal
    cRampDown[t in INTERIORS, g in G],
        vGEN[t-1,g] - vGEN[t,g] <= generators.Ramp_Dn_Percentage[g]*vCAP[g]
    # (11b) Ramp down, sub-period wrapping
    cRampDownWrap[t in STARTS, g in G],
        vGEN[t+hours_per_period-1,g] - vGEN[t,g] <= generators.Ramp_Dn_Percentage[g]*vCAP[g]

    # (12a) Storage state of charge, normal
    cSOC[t in INTERIORS, g in STOR],
        vSOC[t,g] == vSOC[t-1,g] + generators.Eff_Up[g]*vCHARGE[t,g] - vGEN[t,g]/generators.Eff_Down[g]
    # (12a) Storage state of charge, wrapping
    cSOCWrap[t in STARTS, g in STOR],
        vSOC[t,g] == vSOC[t+hours_per_period-1,g] + generators.Eff_Up[g]*vCHARGE[t,g] - vGEN[t,g]/generators.Eff_Down[g]
end)
```

a) Basic scenario

In this scenario, there are no changes, the above model was used as is.

b) 50% decreased population scenario

In this scenario, the Load (MW) data has been changed as shown below to reflect the case of population growth of less than 50% from the basic scenario:

50% decreased popualtion- Load (MW) in case of population growth more than 50% from the basic

```julia
In [4]: # Given values
k = 1.0625  # Proportionality constant
rate_population_growth = 0.016  # Rate of population growth

# Calculate the new rate of population growth when decreased by 50%
new_rate_population_growth_less = 0.5 * rate_population_growth

# Calculate the new rate of power consumption growth using the proportional equation
new_rate_power_consumption_growth = k * new_rate_population_growth_less

# Print the results
println("Original Rate of Population Growth: ", rate_population_growth)
println("New Rate of Population Growth: ", new_rate_population_growth_less)
println("Estimated New Rate of Power Consumption Growth: ", new_rate_power_consumption_growth)
```

Original Rate of Population Growth: 0.016
New Rate of Population Growth: 0.008
Estimated New Rate of Power Consumption Growth: 0.0085

```julia
In [5]: # Calculate modified demand values based on the new rate of population growth decreased by 50%
demand_data[!, :Less_Load_MW_z1] = demand_data[!, :Load_MW_z1] .* (1 + k * (new_rate_population_growth_less - rate_population_gro
demand_data[!, :Less_Load_MW_z2] = demand_data[!, :Load_MW_z2] .* (1 + k * (new_rate_population_growth_less - rate_population_gro
demand_data[!, :Less_Load_MW_z3] = demand_data[!, :Load_MW_z3] .* (1 + k * (new_rate_population_growth_less - rate_population_gro

# Print or display the modified DataFrame
@show demand_data[!, [:Time_Index, :Load_MW_z1, :Load_MW_z2, :Load_MW_z3, :Less_Load_MW_z1, :Less_Load_MW_z2, :Less_Load_MW_z3]]
```

demand_data[!, [:Time_Index, :Load_MW_z1, :Load_MW_z2, :Load_MW_z3, :Less_Load_MW_z1, :Less_Load_MW_z2, :Less_Load_MW_z3]] =
480×7 DataFrame

| Row | Time_Index Int64 | Load_MW_z1 Int64 | Load_MW_z2 Int64 | Load_MW_z3 Int64 | Less_Load_MW_z1 Float64 | Less_Load_MW_z2 Float64 | Less_Load_MW_z3 Float64 |
|---|---|---|---|---|---|---|---|
| 1 | 1 | 350 | 39067 | 1770 | 347.025 | 38734.9 | 1754.96 |
| 2 | 2 | 345 | 38487 | 1744 | 342.067 | 38159.9 | 1729.18 |
| 3 | 3 | 345 | 38554 | 1747 | 342.067 | 38226.3 | 1732.15 |
| 4 | 4 | 348 | 38926 | 1764 | 345.042 | 38595.1 | 1749.01 |
| 5 | 5 | 355 | 39607 | 1795 | 351.983 | 39270.3 | 1779.74 |

c) 50% increased population scenario

In this scenario, the Load (MW) data has been changed as shown below to reflect the case of population growth of more than 50% from the basic scenario:

50% increased popualtion- Load (MW) in case of population growth more than 50% from the basic

```julia
In [2]: # Given values
k = 1.0625  # Proportionality constant
rate_population_growth = 0.016  # Rate of population growth

# Calculate the new rate of population growth when increased by 50%
new_rate_population_growth = 1.5 * rate_population_growth

# Calculate the new rate of power consumption growth using the proportional equation
new_rate_power_consumption_growth = k * new_rate_population_growth

# Print the results
println("Original Rate of Population Growth: ", rate_population_growth)
println("New Rate of Population Growth: ", new_rate_population_growth)
println("Estimated New Rate of Power Consumption Growth: ", new_rate_power_consumption_growth)

Original Rate of Population Growth: 0.016
New Rate of Population Growth: 0.024
Estimated New Rate of Power Consumption Growth: 0.025500000000000002
```

```julia
In [3]: using CSV, DataFrames

# Load data from CSV
demand_data = CSV.read("Load_data.csv", DataFrame)

# Calculate modified demand values based on the new rate of population growth increased by 50%
demand_data[!, :more_Load_MW_z1] = demand_data[!, :Load_MW_z1] .* (1 + k * (new_rate_population_growth - rate_population_growth))
demand_data[!, :more_Load_MW_z2] = demand_data[!, :Load_MW_z2] .* (1 + k * (new_rate_population_growth - rate_population_growth))
demand_data[!, :more_Load_MW_z3] = demand_data[!, :Load_MW_z3] .* (1 + k * (new_rate_population_growth - rate_population_growth))

# Print or display the modified DataFrame
@show demand_data[!, [:Time_Index, :Load_MW_z1, :Load_MW_z2, :Load_MW_z3, :more_Load_MW_z1, :more_Load_MW_z2, :more_Load_MW_z3]]
```

demand_data[!, [:Time_Index, :Load_MW_z1, :Load_MW_z2, :Load_MW_z3, :more_Load_MW_z1, :more_Load_MW_z2, :more_Load_MW_z3]] =
480×7 DataFrame

| Row | Time_Index Int64 | Load_MW_z1 Int64 | Load_MW_z2 Int64 | Load_MW_z3 Int64 | more_Load_MW_z1 Float64 | more_Load_MW_z2 Float64 | more_Load_MW_z3 Float64 |
|---|---|---|---|---|---|---|---|
| 1 | 1 | 350 | 39067 | 1770 | 352.975 | 39399.1 | 1785.04 |
| 2 | 2 | 345 | 38487 | 1744 | 347.933 | 38814.1 | 1758.82 |
| 3 | 3 | 345 | 38554 | 1747 | 347.933 | 38881.7 | 1761.85 |
| 4 | 4 | 348 | 38926 | 1764 | 350.958 | 39256.9 | 1778.99 |
| 5 | 5 | 355 | 39607 | 1795 | 358.017 | 39943.7 | 1810.26 |
| 6 | 6 | 383 | 42767 | 1938 | 386.255 | 43130.5 | 1954.47 |
| 7 | 7 | 424 | 47320 | 2144 | 427.604 | 47722.2 | 2162.22 |
| 8 | 8 | 467 | 52175 | 2364 | 470.969 | 52618.5 | 2384.09 |

2. CO2 limit scenarios with changing population growth rates

Three different scenarios have been developed using the modified load data that reflects different population growth changes and adding the below constraint to all models:

```julia
# (10) CO2 Limit
    sum((generators.CO2_Rate[g] * vGEN[t, g] + generators.CO2_Per_Start[g] * vGEN[t, g])
        for t in T, g in G) <= 0.521 * sum(vGEN[t, g] for t in T, g in G)
```

3. Carbon pricing scenarios

Three different scenarios have been developed Using the modified load data that reflects different population growth changes and adding the below costs to all models:

```julia
# Modify the objective function to include the cost of emissions allowances
@expression(Expansion_Model_price, eVariableCostsWithAllowances,
    eVariableCosts + 22.5 *sum((generators.CO2_Rate[g] * vGEN[t, g] + generators.CO2_Per_Start[g] * vGEN[t, g])
        for t in T, g in G)
)

@objective(Expansion_Model_price, Min,
    eFixedCostsGeneration + eFixedCostsStorage + eFixedCostsTransmission +
    eVariableCostsWithAllowances + eNSECosts
);
```

**Results & Discussion**

Overall Costs

We ran the above models and recorded the total costs for each one to find the solution that minimizes costs with the given constraints and conditions, Table 2 shows the results.

*Table 2 Total Costs Results*

| Scenario | Total Costs ($) |
|---|---|
| Basic _No CO2 Limit/ Price | 6.392e+09 |
| 50% Decreased Population _No CO2 Limit/ Price | 6.333e+09 |
| 50% Increased Population_ No CO2 Limit/ Price | 6.451e+09 |
| Basic_CO2 Limit | 6.392e+09 |
| 50% Decreased Population _CO2 Limit | 6.333e+09 |
| 50% Increased Population_CO2 Limit | 6.451e+09 |
| Basic_ Carbon Pricing | 6.442e+09 |
| 50% Decreased Population _Carbon Pricing | 6.382e+09 |
| 50% Increased Population_ Carbon Pricing | 6.502e+09 |

In the basic scenario without CO2 limits or carbon pricing and without changing load, the total cost is $6.392 billion. Introducing a 50% decrease in population under the same conditions slightly reduces the total cost to $6.333 billion, while a 50% increase in population raises the total cost to $6.451 billion. This observation underscores that alterations in population growth rates significantly impact the overall cost structure.

In scenarios incorporating a CO2 limit of 0.521 tCO2/MWh, the total cost remains the same as the scenarios that include a change of population growth rate only, $6.392 billion for the basic case, whereas a 50% decrease or increase in population results in costs of $6.333 billion and $6.451 billion, respectively. A potential reason for not changing the costs is that the proposed CO2 limit will be already achieved by the year 2030, and adding this constraint will not affect the results.

Introducing carbon pricing with a price of $22.50 per ton of CO2 affected the total cost where the basic scenario with this policy became $6.442 billion and in case of a 50% decrease in population, the total costs slightly decreased to $6.382 billion, while a 50% increase in population raises it to $6.502 billion. These outcomes emphasize the significance of thoroughly studying the implications of implementing a price on CO2 emissions with changing population growth.

The impact of each cost on the overall expenditure for every scenario is illustrated in Figure 7. As indicated in the figure, fixed generator costs emerge as the primary contributors. Figure 8 demonstrates how the transition from one scenario to another influences these costs.

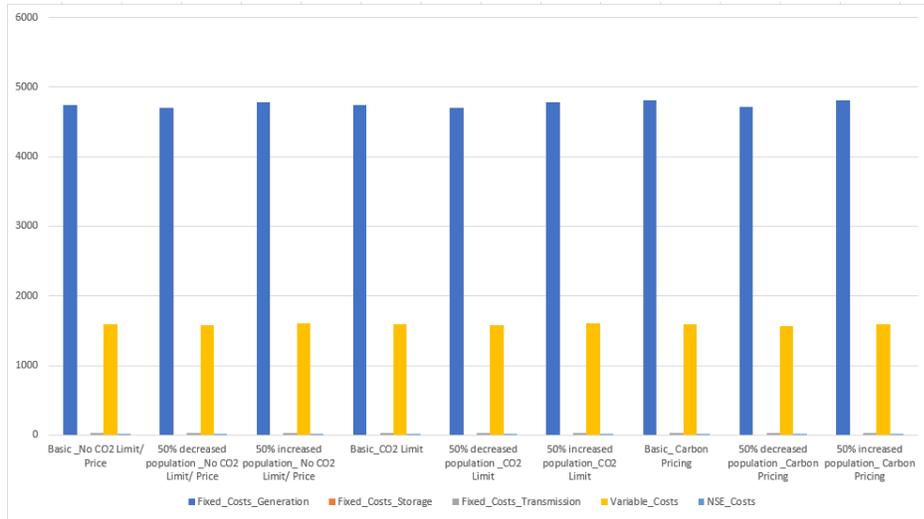
*Figure 7 Different Costs for Each Scenario*

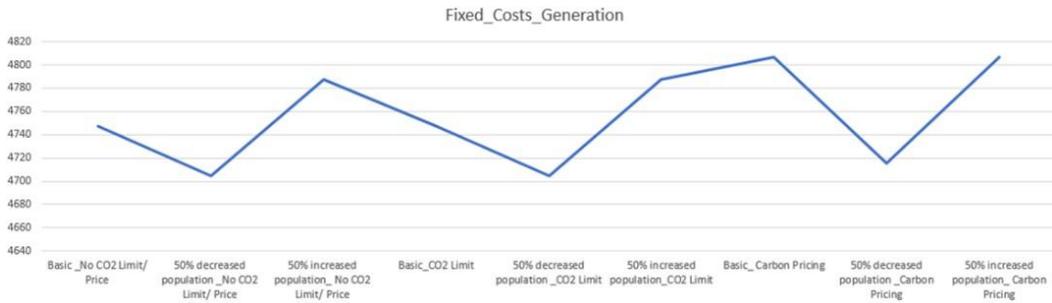
*Figure 8 Fixed Cost Generation for Each Scenario*

Generators Mix Results (MW)

The impact of changes in population growth and the implementation of decarbonization policies, including CO2 limits and carbon pricing, on generation by resource type is depicted in Figure 9. When comparing variations across different population rates, it is observed that the percentage differences among the three scenarios exhibit minimal fluctuations, only decreasing or increasing by approximately 1% with changes in population rates. Solar represents the largest share of generation, followed by wind, natural gas, and lastly, batteries and Trans light, which maintain nearly identical percentages across all scenarios.

For the effect of the policies, applying CO2 limits doesn't affect the distribution of generators, while applying carbon pricing does. When implemented, it is evident that the percentages of clean energy, including solar and wind, increase and decrease for natural gas across different population rates. This suggests that putting a price on carbon emissions can contribute to lowering them and aid in achieving the goal of establishing reliable and sustainable energy sources.

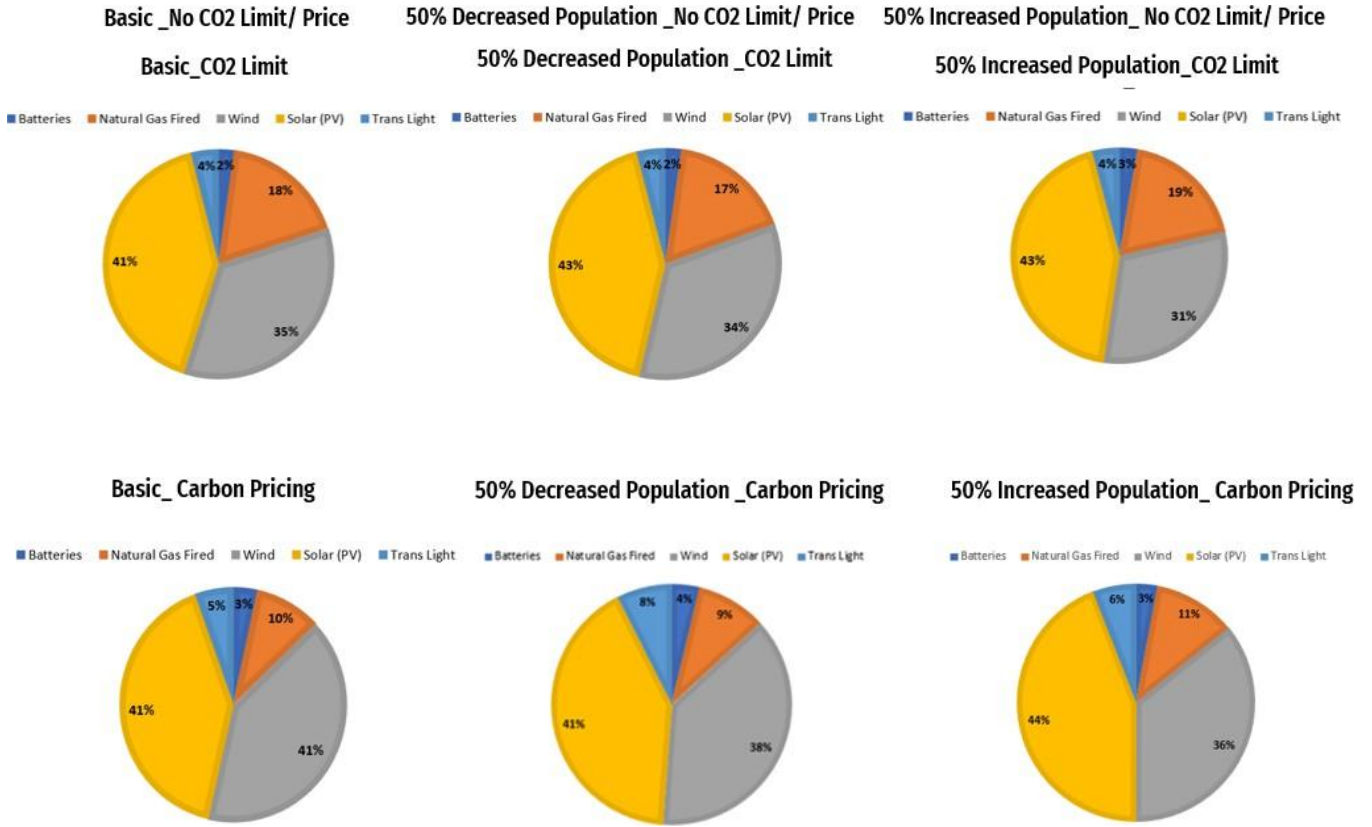

*Figure 9: Different Generator Distribution for Different Scenarios*

Non-Served Energy (NSE) (MWh)

Figure 10 shows the NSE for each scenario. As the population increases, the NSE for all three considered zones also increases, with slight differences between one rate and another. The addition of the carbon pricing policy contributes to an increase in NSE, potentially because it introduces economic incentives for cleaner energy practices, leading to a shift towards more sustainable and efficient energy generation.

**Basic_CO2 Limit**
**Basic _No CO2 Limit/ Price**

| Zone | Total_NSE_MWh |
|---|---|
| 1 | 348.15 |
| 2 | 26588.72277 |
| 3 | 1763.025 |

**50% Decreased Population _CO2 Limit**
**50% Decreased Population _No CO2 Limit/ Price**

| Zone | Total_NSE_MWh |
|---|---|
| 1 | 345.190725 |
| 2 | 26378.94853 |
| 3 | 1748.039288 |

**50% Increased Population_CO2 Limit**
**50% Increased Population_ No CO2 Limit/ Price**

| Zone | Total_NSE_MWh |
|---|---|
| 1 | 351.109275 |
| 2 | 26799.9678 |
| 3 | 1778.010712 |

**Basic_ Carbon Pricing**

| Zone | Total_NSE_MWh |
|---|---|
| 1 | 350.1093 |
| 2 | 27800.43 |
| 3 | 1778.011 |

**50% Decreased Population _Carbon Pricing**

| Zone | Total_NSE_MWh |
|---|---|
| 1 | 347.1907 |
| 2 | 26380.04 |
| 3 | 1760.039 |

**50% Increased Population_ Carbon Pricing**

| Zone | Total_NSE_MWh |
|---|---|
| 1 | 356.1093 |
| 2 | 26820.43 |
| 3 | 1780.011 |

*Figure 10: NSE (MWh) for Different Scenarios*

Total Storage and Transfer Results (MWh)

The results indicate that the total storage resources for each zone and the transfer capacity remain unchanged across all scenarios. This suggests that population variations and the implemented decarbonization policies do not have a discernible effect on these parameters. Figure 11 represents these consistent outcomes.

**Total Storage Results**

| Zone | Resource | Total_Storage_MWh |
|---|---|---|
| 1 | ERC_PHDL_batteries_1 | 1600 |
| 2 | ERC_REST_batteries_1 | 13006.9 |
| 3 | ERC_WEST_batteries_1 | 9328.9 |

**Total Transfer Results**

| Line | Total_Transfer_Capacity | Start_Transfer_Capacity | Change_in_Transfer_Capacity |
|---|---|---|---|
| 1 | 5664 | 3332 | 2332 |
| 2 | 16529 | 16529 | 0 |

*Figure 11 Total Storage and Transfer Results (MWh)*

**Conclusions**

The study reveals that population growth in the ERCOT area has a direct and discernible impact on the energy market, influencing overall costs and NSE results. As the population increases, it leads to a corresponding escalation in both costs and NSE. Interestingly, this population growth does not affect total storage and total transfer results, suggesting a level of stability in these aspects regardless of demographic changes. For the generation mix, solar remains the predominant source of generation, followed by wind, natural gas, and batteries/Trans light with very slight change between different population growth rates.

Additionally, the study identified that the introduction of a $CO_2$ limit has a negligible impact on the ERCOT energy market. This outcome may be attributed to the observation that Texas has already achieved the proposed $CO_2$ limit for the year 2030. Consequently, incorporating this constraint does not yield discernible changes in the results. Accordingly, the outcomes for scenarios involving $CO_2$ limits remain consistent with those of the basic scenarios, only varying with changes in population growth rates.

In contrast, the introduction of carbon pricing proves influential, resulting in increased overall costs across all population scenarios. Moreover, carbon pricing promotes the utilization of clean energy resources and contributes to enhanced NSE results. The observed increase in clean energy percentages, coupled with a decrease in natural gas, emphasizes the effectiveness of carbon pricing in steering toward sustainable energy practices.

For potential future work, extending the model to incorporate emerging technologies and conducting comparative studies on other interconnection areas would provide valuable insights into the evolving dynamics of energy markets and decarbonization policy implementations.